\documentclass[pra,aps,twocolumn,showpacs,preprintnumbers]{revtex4}

\usepackage{epsfig}
\usepackage{amsmath}
\usepackage{graphicx}
\usepackage{amsmath}
\usepackage{array,color}
\begin{document}

\title{Arbitrary-order lensless ghost imaging with thermal light}
\author{Xi-Hao Chen$^1$, Ivan N. Agafonov$^2$, Kai-Hong Luo$^1$,
Qian Liu$^1$, Rui Xian$^1$, Maria V. Chekhova$^2$, and Ling-An
Wu$^1$}
\thanks{Corresponding author: wula@aphy.iphy.ac.cn}
\affiliation{$^1$Laboratory of Optical Physics, Institute of Physics
and Beijing National Laboratory for Condensed Matter Physics,
Chinese Academy of Sciences, Beijing 100190, China
\\\it{$^2$} Department of Physics, M. V. Lomonosov Moscow State
University, Leninskie Gory, 119992 Moscow, Russia}

\begin{abstract}
Arbitrary $N$th-order ($N\geq2$) lensless ghost imaging with thermal
light has been performed for the first time by only recording the
intensities in two optical paths. It is shown that the image
visibility can be dramatically enhanced as the order $N$ increases.
It is also found that longer integration times are required for
higher-order correlation measurements as $N$ increases, due to the
increased fluctuations of higher-order intensity correlation
functions.

\end{abstract}


\maketitle

Compared with the first ``ghost'' imaging experiment with two-photon entangled
light~\cite{Pittman}, second-order ghost imaging and ghost interference with
thermal
light~\cite{Gatti,Ferri,Valencia,Zhu,wang,Zhangda,Zhai,Zhai1,Scarcelli1,Chen}
has a low visibility which theoretically can never exceed 1/3, and in fact will
be much lower than 1/3 in practical applications. Moreover, when a
2-dimensional high resolution image of a complex object is required, the better
the resolution, the worse will its visibility be. This is one of the
limitations for the practical application of thermal ghost imaging (GI).
Fortunately, recent studies~\cite{Liu,Han,Agafonov,Cao,Richter,Agafonov1} on
the higher-order intensity correlation effects of thermal light show that the
visibility can be significantly improved by increasing the order $N$. In this
way, the drawback of low visibility in correlated imaging with thermal light
can be overcome.

Third-order GI with thermal light has been theoretically analyzed to
a certain extent ~\cite{Kuang,Han}, but recently Liu \textit{et al}
pointed out that it is  inappropriate to assume that second-order
correlations play the entire or dominant role \cite{Liu}. In their
investigations of higher-order thermal ghost imaging and
interference Liu \textit{et al} showed that it is $N$-photon
bunching that characterizes the $N$th-order correlation and leads to
the high-visibility in $N$th-order schemes. The necessary condition
for achieving a ghost image or interference pattern in $N$th-order
intensity correlation measurements is the synchronous detection of
the same light field by different reference detectors. Multi-photon
interference experiments have been carried out by Agafonov
\textit{et al} ~\cite{Agafonov}, verifying the conclusion that the
visibility limits of three-photon and four-photon interference are
respectively $82\%$ and $94\%$ for classical coherent light, as
predicted theoretically by Richter~\cite{Richter}. Cao \textit{et
al} discussed $N$-th order intensity correlation in double-slit
ghost interference with thermal light and proposed a scheme to study
the visibility and resolution of the fringes with two
detectors~\cite{Cao}. However, in their actual experiment only one
CCD detector was employed, and the measurements were taken first
with and then without the double-slit in place. Similar  high-order
schemes to obtain higher visibility but for GI were also suggested
by Agafonov \textit{et al} a little earlier~\cite{Agafonov1}.

In this paper we report the first demonstration of an arbitrarily
high $N$th-order lensless GI experiment with pseudothermal
radiation. We do not actually need $N$ light paths but measure the
$N$th-order intensity correlations by means of just two detectors.
Moreover, when certain conditions are met, no lens is required for
obtaining a well-focused image of the object.

Generally, in an $N$th-order intensity correlation measurement, the
light beam needs to be divided into $N$ parts, each of which passes
through an optical system and then is registered by a detector. In
the high-intensity limit, the normalized $N$th-order intensity
correlation function is given by~\cite{Cao,Mandel}

\begin{eqnarray}\label{gN}
g^N (x_1 ,...,x_N ) &=& \frac{{\left\langle {I_1 (x_1 )...I_N (x_N
)} \right\rangle }}{{\left\langle {I_1 (x_1 )} \right\rangle
...\left\langle {I_N (x_N )} \right\rangle }},
\end{eqnarray}
where $I_j(x_j)$ is the instantaneous intensity at position $x_j$ in the
transverse direction, and $\langle...\rangle$ stands for ensemble averaging, in
practice achieved through time averaging. It is well known that
$g^N(x)=\left\langle I^N(x) \right\rangle/ \left\langle I(x) \right\rangle^N
=N!$ for polarized thermal light when all the points in Eq.~\ref{gN} are the
same space-time point~\cite{Mandel,Loudon}. This $N$-th order intensity
correlation function $g^N$ can thus be obtained by measuring the
autocorrelation function of $I(x)$, as has been done in the 2nd order GI and
interference-diffraction experiments with pseudothermal light
\cite{Ferri,Basano,Cao}, with mere knowledge of the intensity distribution
$I(x)$ of that space-time point. It is thus reasonable to envisage a setup in
which we have $n$ beams of one intensity distribution (say passing through an
object) and $N-n$ beams of another distribution (as reference arms), so that
the instantaneous intensities $I_{1}(x_1)=I_{2}(x_2)=...=I_{n}(x_n)=I(y_1)$ and
$I_{n+1}(x_{n+1})=I_{n+2}(x_{n+2})=...=I_{N-n}(x_{N-n})=I(y_2)$. Then the
transverse normalized $N$th-order correlation function may be defined as
 \begin{eqnarray}  \label{g2}
 g^{(N)}_{n}(y_1,y_2) =\frac{\langle I^{n}(y_1)I^{N-n}(y_2)\rangle}
 {\langle I(y_1)\rangle^{n}\langle I(y_2)\rangle^{N-n}},
\end{eqnarray}
where the intensity correlation is composed of an $n$-fold intensity
product at position $y_1$ and an ($N-n$)-fold product at position
$y_2$, so there are $N-1$ ways to measure the $N$th-order
correlation function~\cite{Liu,Cao}. This equation implies that, in
GI, an arbitrary order correlation function can be obtained with
mere knowledge of the two intensity distributions $I(y_1)$ and
$I(y_2)$.

Here we consider a simple experimental scheme such as that shown in
Fig. 1, which is basically the same as that for lensless 2nd-order
correlation ghost imaging~\cite{Chen}. To pass to higher-order
intensity correlation measurements, the signals of the test and
reference detectors must be raised to the powers $n$ and $N-n$. It
should be mentioned that the correlation function measured this way
is not the same as that in an $N$-port HBT interferometer, but the
difference vanishes in the high-intensity limit as classical
intensity fluctuations (excess noise) become stronger than quantum
fluctuations. We consider the intensity distributions $I(y_1)$ and
$I(y_2)$ of the speckle fields on the $Z_1$ and $Z_2$ planes
located, respectively, at distances $z_1$ and $z_2$ from the thermal
source. For simplicity but without loss of generality, in the
derivation below the image is considered to be one-dimensional in
the $Z_1$ plane.

 If $T(y)$ indicates the field transmission function of the object,
and a bucket detector is used in the test (object) arm, the bucket
signal $S$ is
\begin{eqnarray}  \label{S}
S = \int {I(y_1 )\left| {T(y_1 )} \right|} ^2 dy_1 .
\end{eqnarray}
If we assume uniform illumination so that $ \langle I(y_1
)\rangle=I$, the average value of $S$ is
\begin{eqnarray}  \label{S}
\langle S \rangle = I\int {\left| {T(y_1 )} \right|^2 dy_1 = I}
A_{obj} ,
\end{eqnarray}
in which $A_{obj}$ is equal to the object area when $\left| {T(y_1
)} \right|^2$ is a step function equal to 1 inside the object and 0
outside.
\begin{figure}[t]
\includegraphics[width=7.5cm]{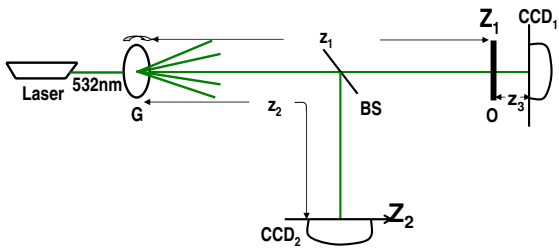}\caption{(Color online)
Experimental setup for arbitrary-order lensless GI with a
pseudothermal light source. G: rotating ground glass plate; BS:
beamsplitter; O: object.} \label{setup}
\end{figure}
In the classical limit we can also write the $N$th-order correlation
function for GI as
\begin{eqnarray}\label{gn}
\gamma_n ^{(N)} (y_2 ) &=& \frac{{\left\langle {S^n I^{N - n} (y_2
)} \right\rangle }}{{\left\langle S \right\rangle ^n \left\langle
{I(y_2 )} \right\rangle ^{N - n} }}
\end{eqnarray}
where we have assumed that all the $n$ test arms have the same configuration,
i.e., contain the same objects in the same positions and identical bucket
detectors. It is evident that this requirement can be simply satisfied with a
single test beam and the detector output divided into $n$ parts. Since the same
reasoning can be applied to the reference arms, we can thus perform any
$N$th-order correlation measurement with a simple experimental setup similar to
that for 2nd-order lensless imaging in which only two light paths are needed,
as shown in Fig.~\ref{setup}. The reference detector output is correspondingly
divided into $N-n$ parts~\cite{Chan}. From the viewpoint of photodetection, the
relation between the photocurrent $i(x,t)$ and light intensity $I(x,t)$
is~\cite{Mandel} $i(x,t)\propto (1/T)\int_{ - T/2}^{T/2} {I(x,t + \tau )d\tau }
$, where $T$ is the exposure time of the detector. If $s$ and $i$ are the
photocurrents of CCD1 and CCD2, respectively, then from Eq.~\ref{gn} the
$N$th-order correlation function for GI is
\begin{eqnarray}\label{r}
\gamma_n ^{(N)} (y_2 ) &=& \frac{{\left\langle {(s/n)^n
(i(y_2)/(N-n))^{N - n}} \right\rangle }}{{\left\langle s/n
\right\rangle ^n \left\langle {i(y_2 )/(N-n)} \right\rangle ^{N - n}
}}.
\end{eqnarray}
This is the expression  that is actually used in our data
processing. The total intensity of the light source is assumed to be
equal to the sum of the intensities of all the test and reference
paths.

\begin{figure}[t]
\includegraphics[width=7cm]{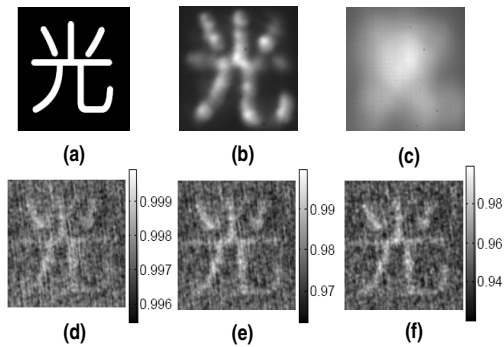}
\caption{First row: (a) Object mask. (b) and (c): Projection images
obtained by CCD1 alone, averaged over 20,000 frames, for (b) $z_3$ =
20 mm and (c) $z_3$ = 70 mm. Second row: High-order ghost images
obtained with thermal light in a lensless setup, averaged over
140,000 frames, corresponding to (d) 2nd order ($N=2, n=1$), (e)
10th order ($N=10, n=9$), (f) 20th order ($N=20, n=19$).}
\label{image-1}
\end{figure}

 An outline of the experimental set-up is shown in Fig.~\ref{setup}.
 A CW laser beam with a wavelength $\lambda$ of 532~nm
and beam diameter $D$ of 3~mm is projected onto a ground-glass plate G rotating
at a speed of 0.7 rad/s to form pseudothermal light. The scattered beam is
separated by a $50\%-50\%$ non-polarizing beamsplitter (BS) into two beams. The
transmitted beam passes through the object, and the reflected beam is the
reference beam. The two beams are detected by the charged-coupled-device (CCD)
cameras CCD1 (Imaging Source DMK 21BU04) and CCD2 (Imaging Source DMK 31BU03),
respectively. The object (a mask), which is the Chinese character for ``light''
shown in Fig.~\ref{image-1}(a), is placed in the test arm where CCD1 plays the
role of a bucket detector. The distance $z_1$ between the source and the object
is equal to the distance between the source $z_2$ and CCD2, namely,
$z_1$=$z_2$=240~mm. Thus the coherence area $A$ at the plane $Z_1$ is about 16
$\mu$m$^2$, according to the well-known relationship of $ A \sim ({\lambda
z_1}/D)^2 $; $z_3$ is the distance between the mask and CCD1 and is equal to 70
mm for the high-order imaging experiments. The cameras are operated in the
trigger mode and are synchronized by the same trigger pulse. The data are
acquired with an exposure time (0.1~ms) much shorter than the correlation time
of the light source, which is on the order of 0.2~s, and saved through a USB
cable to a computer.

Figures 2(b) and (c) show the images of the mask obtained by direct
exposure on the camera CCD1 alone for $z_3$ = 20~mm and 70~mm (not
in the bucket detection mode), after averaging over 20,000 frames.
We see that Fig.~2(c) is a completely blurred image due to
interference-diffraction effects, while Fig.~2(b) is quite a clear
projected image since the CCD camera is close behind the mask.

Though we cannot obtain a first-order image in the case of $z_3$
=70~mm, we can obtain clearly discernible higher-order ghost images
($N\geq 2$), as shown in the lower half of Fig.~2. For these shots a
total of 140,000 frames was grabbed by each camera per plot at a
rate of about 1.35 Hz. The arbitrary-order correlation images are
calculated according to Eq.~\ref{r} by a Matlab program in which the
total intensity $s$ of each frame recorded by the bucket detector
CCD1 is divided by $n$ to give the intensity $s/n$ of each component
test arm, and similarly the output $i(y_2)$ of CCD2 which is
captured at the same instant is divided by $N-n$ to give the
intensity of each component reference beam; then the product of
$(s/n)^{n}$ and ${(i(y_2)/(N-n)})^{N-n}$ is averaged over all the
exposure shots and divided by ${{\left\langle s/n \right\rangle ^n
\left\langle {i(y_2 )/(N-n)} \right\rangle ^{N - n} }}$ to give the
normalized $N$th-order correlation function $\gamma_n ^{(N)} (y_2
)$. Here $\left\langle s \right\rangle $ and $\left\langle {i(y_2 )}
\right\rangle $ are obtained by averaging all the frames of CCD1 and
CCD2, respectively. In our experiment $n=N-1$ is chosen, as this
gives the best visibility with the minimum number of frames to be
processed. The reconstructed 2nd, 10th and 20th order ghost images
which are normalized following $ {\gamma_n ^{(N)} (y_2 )}/{(\gamma_n
^{(N)} (y_2 ))_{Max} }$ are shown in Figs. 2(d), (e) and (f),
respectively. It can be clearly seen that the image visibility
increases with increasing $N$, as predicted. This is one of the
chief advantages of GI systems. Another advantage is that the
position distribution of the light intensity transmitted through the
mask is not necessary in any-order GI, so long as all of the light
is captured by the bucket detector. This has been demonstrated by
Meyers \textit{et al} who succeeded in obtaining 2nd-order ghost
images even with a turbid distortive medium placed between the
object and bucket detector~\cite{Meyers}.

It should be remarked that the rate of increase of the visibility in
object imaging is much lower than that in interference experiments,
such as the $N$th-order Hanbury Brown-Twiss and interference
experiments reported by Cao \textit{et al}~\cite{Cao}. This is
because the number $M_{obj}=A_{obj}/A_{coh}$ of coherent areas
falling inside the object, which is defined in Ref.~\cite{Ferri1},
is very high and the visibility decreases with the increase of
$M_{obj}$. Moreover, it should be noted that there is, in fact, a
different number of coherent areas $M_{obj}^{(N)}$ within the object
for each different order intensity correlation measurement if we
define the coherent area (the resolvable area under the focused
condition) as the square of the full-width-half-max $\delta y^{(N)}$
of the $N$th-order intensity correlation function. For a given $N$,
$M_{n,obj}^{(N)}$ is also different when $n$ is different. It has
been shown that the visibility increases with the order $N$ when $N$
is not very high, while it is independent of the resolution and
close to unity at very high $N$~\cite{Cao,Agafonov1}.

\begin{figure}[t]
\includegraphics[width=6cm]{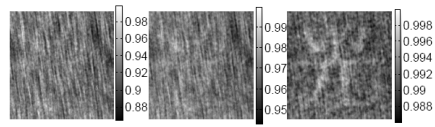}
\caption{4th-order lensless ghost images for $n= 1, 2$, and 3 (from
left to right).} \label{image-2}
\end{figure}
\begin{figure}[b]
\includegraphics[width=7cm]{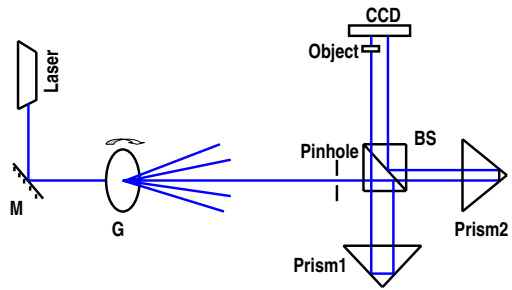}
\caption{(Color online) Experimental setup for arbitrary-order
lensless GI with a pseudothermal light source. G: rotating ground
glass plate; BS: beamsplitter; M: mirror.} \label{setup-2}
\end{figure}

It should be pointed out that in $N$th-order lensless GI there are
at least $N-1$ experimental configurations, since $n$ may be
different. We take the 4th-order as an example to show the
difference between them. In Fig.~3 the 4th-order ghost images for
$n=1, 2$, and 3 are obtained by processing the same number of frames
(i.e. using the same integration time) after the exposures with the
CCD cameras. It is evident that the image is completely blurred in
the case of $n=1$ while a good image is obtained when $n=3$. It is
found that the higher the order $N$, the longer is the integration
time required for a clear image to be obtained. Moreover, for a
given $N$ the lower the $n$, the longer is the integration time
required. This can be inferred from Eq.~\ref{r} where we can see
from the self-correlation intensity product $i^{N-n}(y_2)$ between
the reference signals (or $s^n$ for the test signals) that as $n$
decreases (or $N-n$ increases), the fluctuations of the higher-order
correlation functions will increase. Since the fluctuation of the
total intensity recorded by the bucket detector (the $N$th-order
self-correlation function of $s$) is less than the sum of the
fluctuations of each pixel in the reference detector ($i{(y_2)}$),
as we have indeed observed experimentally, it is best to choose a
large $n$ to obtain clearer images and save integration time.

\begin{figure}[t]
\includegraphics[width=7cm]{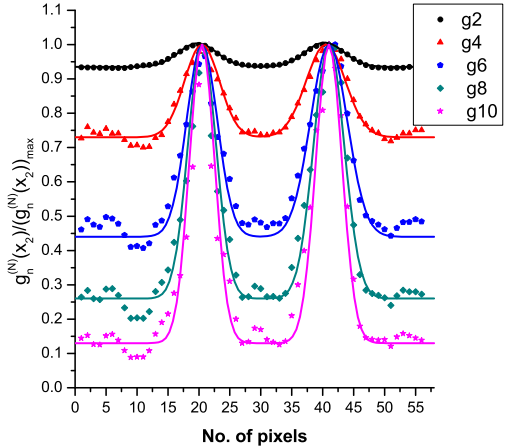}
\caption{(Color online) Experimental results of 2nd, 4th, 6th, 8th
and 10th order lensless GI along one cross-section. Transverse
coordinates are the number of pixels, and each point on the plot is
an average of the raw data from 5 pixels. The solid lines are
Gaussian fits.} \label{slit-image}
\end{figure}
We have also performed an experiment for a smaller $M_{obj}$ using
the experimental setup shown in Fig.~\ref{setup-2}. A He-Cd laser of
wavelength 441.6~nm is reflected by a mirror (M) and used with a
rotating ground glass plate (G) to form the pseudothermal light
source, which is about 1~mm in diameter. A pinhole is placed 18~cm
away from the ground-glass plate to limit the transverse size of the
beam, which is separated by a $50\%-50\%$ non-polarizing
beamsplitter (BS) into two beams. The reflected test beam is totally
internally reflected by prism1, passes through BS and the object,
and is detected on the lefthand side of a CCD camera (Imaging Source
DMK 31BU03). The transmitted reference beam is reflected by prism2
and then by BS to be recorded on the righthand side of the detector.
Both prisms can be translated to adjust the beam path distances
between the light source and the detector. The object is a
double-slit mask with slits of width 150~$\mu$m and separation
570~$\mu$m. As before, for lensless imaging the distances from the
source to the object and reference detector are equal,
$z_1$=$z_2$=354 mm. To make $M_{obj}$ small enough, the data is only
collected in one dimension along the transverse direction. The
experimental results are shown in Fig.~\ref{slit-image}, in which
$n=N/2$, and the 2nd, 4th, 6th, 8th and 10th order image
cross-sections are obtained by averaging about 4,000 exposure
frames. We can see that the image visibility increases much faster
than in our previous experiment, mainly because here the number of
features $M_{obj}$ is much smaller.

In conclusion, we have reported the first demonstration of
arbitrary-order ($N\geq 2$) lensless GI with pseudothermal light by
only recording the intensities in two optical paths. The
experimental results demonstrate that the image visibility can be
dramatically enhanced as the order $N$ increases. This overcomes the
bottleneck of low visibility in 2nd-order ghost imaging with thermal
radiation. It is conceivable that higher order GI could also be
realized with a single detector setup using a laser and computer
generated spatial modulation~\cite{Shapiro, Bromberg}, but for a
real thermal light source the field distribution in the reference
arm must be actually measured by at least one nonvirtual detector.
As we have shown, in fact just one reference detector is sufficient.
It is also found that longer integration times are needed as $N$
increases, due to the increased fluctuations of higher-order
intensity correlation functions. Moreover, for a given $N$ the
integration time increases with the decrease of $n$, so it is best
to choose a large $n$ to save time. The experimental setup for
high-order GI is the same as that for the 2nd-order experiment, and
we only require a program to calculate the image to any desired
order of correlation. This makes GI with a thermal source even more
promising than before for practical applications.

This work was supported by the National Natural Science Foundation
of China (NNSFC Grants 60578029 and 10674174), the National Program
for Basic Research in China (Grant 2006CB921107), the Russian
Foundation for Basic Research (RFBR Grant 08-02-00555), and by a
joint grant (RFBR 06-02-39015 GFEN) of RFBR and NNSFC.

\end{document}